# Evidence for a Structurally-driven Insulator-to-metal Transition in VO$_2$: a View from the Ultrafast Timescale.

A.Cavalleri[(1)*], Th. Dekorsy[(2)], H.H. Chong[(1)], J.C. Kieffer[(3)], R.W. Schoenlein[(1)].

[(1)]*Materials Sciences Division, Lawrence Berkeley National Laboratory, Berkeley, CA*
[(2)]*Forschungszentrum Rossendorf, Dresden Germany.*
[(3)]*Université du Québec, INRS énergie et matériaux, Varennes, Québec.*

Abstract

We apply ultrafast spectroscopy to establish a time-domain hierarchy between structural and electronic effects in a strongly-correlated electron system. We discuss the case of the model system VO$_2$, a prototypical non-magnetic compound that exhibits cell doubling, charge localization and a metal-insulator transition below 340 K. We initiate the formation of the metallic phase by prompt hole photo-doping into the valence band of the low-T insulator. The I-M transition is however delayed with respect to hole injection, exhibiting a bottleneck timescale that corresponds to half period of the phonon connecting the two crystallographic phases. This experiment indicates that this controversial insulator may have important band-like character.

Correlated electron materials exhibit remarkable effects, ranging from metal-insulator transitions to non-conventional (High-Temperature) superconductivity. The subtle interplay between atomic structure, charge, spin and orbital dynamics is responsible for many of the critical phenomena observed[1]. Importantly, because "simultaneous" changes in more than one degree of freedom are often observed as chemical doping or external parameters are tuned across critical values, time-integrated spectroscopies are unable to uniquely assign cause-effect relationships.

Here, we demonstrate that time-resolved spectroscopy can instead be applied to overcome such ambiguities. We study the case of non-magnetic $VO_2$, a highly controversial, strongly-correlated compound that exhibits cell doubling in "concomitance" with electron localization and a metal-insulator transition below 340 K[2] (see figure 1). The issue is whether the insulating behavior in the low-T phase derives directly from the Peierls distortion[3] or from electron localization and consequent increase in electron-electron repulsion[4,5]. Recently, a theoretical study by Wentzcovitch et al. has revived attention into this four-decade long debate[6], suggesting that the former mechanism may be dominant, i.e. the low-T phase may be band-like and the transition structurally driven. New controversy has resulted[7,8] and the problem is yet to be settled experimentally.

Previous time-resolved optical[9] and x-ray diffraction[10] experiments in this compound demonstrated that impulsive photo-excitation of the low-T, monoclinic insulator causes an ultrafast transition in both electronic properties *and* atomic structural arrangement. However, it was not clear whether the system becomes metallic due to the change in symmetry of the unit cell or to the prompt creation of holes, causing the closure of a Mott gap. We now report evidence of a limiting structural timescale for the formation of the metallic phase, despite much faster hole-doping into the correlated d band. Such bottleneck time is approximately equal to half period of impulsively excited, coherent optical-phonon distortions that map onto the crystallographic arrangement of the high-T phase. Such evidence for a structurally-mediated transition is suggestive of important band-insulating character of monoclinic $VO_2$. This or analogous experimental strategies may have important and wide applicability to further our understanding of correlation effects in complex solids.

Thin films (50 nm ± 10 nm) of Vanadium Dioxide on Si (111) wafers, with a (200 nm ± 10 nm) Silicon Nitride buffer layer were used for the experiments. Time-resolved optical spectroscopy was performed at several wavelengths and as a function of pulse duration, using a 1-KHz amplified Ti-Sapphire laser system, white-light continuum generation and an optical parametric amplifier. We first conducted 100-fs resolution measurements in transmission and reflection on free-standing $Si_3N_4/VO_2$ structures obtained by chemically etching the Silicon substrate. The normal-incidence 790-nm reflectivity/transmission changes, induced by excitation with 50 mJ/cm$^2$, 100-fs pulses at the same wavelength are reported in figure 2. A sub-picosecond insulator-to-metal transition is evidenced by the abrupt change in the optical properties of the system, resulting in an increase of the reflectivity and a decrease in the transmission. The observed reflectivity/transmission changes persist for tens of nanoseconds and correspond to changes in the complex refractive index (calculated by inverting Fresnel's equation for a two-layer structure), between the equilibrium low-T (n=2.9, k=0.5) and High-T (n=2.3, k=0.72) phases. Experiments at 500-nm, 540-nm, 620-nm and 790-nm lead consistently to the same conclusion. Finally, the observed reflectivity/transmission changes saturate above 25 mJ/cm$^2$, indicative of a complete phase transformation of the film (the film thickness is a factor of two smaller than the 1/e absorption depth). In addition to the unique matching of the refractive index at 1-ps time delay with that of the high-T phase, it is important to point out that the observed response cannot be explained by mere excitation of carriers across a semiconducting bandgap. In fact, e-h pairs at this density would result in a decrease of the reflectivity as opposed to the observed increase. Secondly, the response would exhibit no threshold and would not saturate with fluence. Thirdly, the observed lack of relaxation is not consistent with the behavior of hot carriers. The observed response originates from a non-thermal transition to the metallic phase within less than 1 ps, followed by thermalization of the system in the high-T phase, which then relaxes thermally into the low-T semiconductor by thermal diffusion and nucleation (tens to hundreds of nanoseconds).

The ambiguity on the origin of the photo-induced transition results directly from the strongly correlated nature of this compound. The distorted, low-T crystallographic structure can be derived from of the high-T rutile phase by pairing and tilting of V atoms along the c-axis (Figure 1). The electronic structure of the two phases of VO$_2$ is described along the lines of the Goodenough model[11]. Cell doubling and pairing of the V atoms in the low-T phase splits the half filled d// band by an amount theoretically estimated to be of order 500 meV[12]. Also, deformation of the octahedrally-coordinated oxygen crystal field results from the tilting motion of the V ions, raising the hybridized π$^*$ band above the Fermi level[13,14]. Finally, electron localization on the Vanadium pairs enhances on-dimer Coulomb repulsion (Hubbard U) to approximately 2.0 eV[15], suggestive of a large electronic contribution to the d// splitting.

Photo-excitation of the low-T phase using the photon energies of our experiment corresponds to a transition between the uppermost-occupied 3d band and the hybridized $\pi^*$ band (0.7 eV < hν < 2.5 eV)[16]. Over the range of excitation fluence where the photo-induced phase transition is observed experimentally, the number of absorbed photons per unit volume ranges from 20% to 100% of the valence-band d electrons. Even assuming a major role played by two-photon and excited state absorption events, our experiments span excitation regimes well in excess of half-doping. Therefore, in the case of an insulating state arising primarily from electronic correlations, the collapse of the bandgap should be prompt, without necessity of relaxing the low-T distortion. Alternatively, for a band-like insulator, the timescale for the metallic transition would be set by atomic motion.

In order to address this question, the transition time was measured as a function of pulse duration in the range between 1.5 ps and 15 fs. A non-collinear, optical parametric amplifier and prism compression were used to generate 15-fs pulses, with sufficient energy (~ μJ) to drive the phase transition[17]. Figure 3 shows selected reflectivity responses in a non-etched structure, along with a plot of the transition time as a function of pulse duration, measured at the sample position. In the non-etched structure, the photo-induced transition was evidenced by a decrease in the reflectivity, as opposed to an increase for the free-standing films. This sign reversal is due to the different thickness of the $Si_3N_4$ film and to the modified phase-shift experienced by the light upon reflection at the $Si_3N_4$/Si interface. Similarly to the response of the etched structures, the optical response exhibited a threshold, saturation behavior and could be reconciled with the expected changes in the refractive index across the phase transition. The transition rate was observed to become progressively faster with pulse-duration down to 80 fs pulses, below which a limiting timescale appeared (figure 3).

The observed bottleneck is too fast to be attributed to lattice temperature effects, because in the sub-100 fs timescale electronic excitation is still largely decoupled from the lattice. Coherently initiated structural motion, brought about by optical phonons, is the most likely explanation for the collapse of the bandgap[18,19]. This interpretation is supported by the Raman response of $VO_2$, which was measured in continuous-wave geometry and compared to the time-resolved reflectivity response at low-fluence femtosecond laser excitation. Approximately 10 µJ/cm² were used to excite the $VO_2$ samples, i.e. three orders of magnitude lower than needed to drive the phase transition. Because of the perturbative excitation, vibrational coherence was preserved during relaxation, as evidenced by long-lived, cosine-like oscillations (figure 4a). The Fourier-transform of the time-domain trace is plotted in figure 4b with the continuous-wave Raman spectra. The comparison reveals that only totally symmetric modes of Ag symmetry are impulsively excited, indicative of displacive excitation of coherent phonons[20]. The excited vibrations were observed to disappear when the static temperature was raised above the transition temperature, in agreement with cw Raman behavior. Remarkably, the bottleneck timescale for the phase transition observed at higher fluence corresponds approximately to half period of the two coherent modes.

Among the 18 non-degenerate Raman-active modes of the low-T, $C^5_{2h}$ phase (9 of Ag and 9 of Bg symmetry) we consider the two closely-spaced normal modes that map onto the Rutile structure, with symmetry that is compatible with their disappearance in the high-temperature phase. The structural pathway connecting the two phases has been discussed for the temperature-driven reverse process[21]. Two zone-edge acoustic phonons in the Γ–R (101) direction of the $V_2O_4$ high temperature $D^{14}_{4h}$ phase were identified as the order parameter. These modes are associated with pairing and tilting motions of the high-T rutile metal, have been shown by diffuse x-ray scattering experiments to undergo significant softening on crossing $T_c$[22], and their frequencies have been calculated to be of order 2-4 Thz in the high-T phase[23]. Doubling into the $V_4O_8$ low-T unit cell, folds these phonons to the center of the Brillouin zone, resulting in Ag symmetry and Raman activity. Finally, pairing of the V atoms and electron localization may well stiffen the bonds and result in a 5-6 Thz optical phonon. A rigorous theoretical treatment involving lattice dynamic calculations and assessment of real-space atomic motion is necessary to strengthen this assignment.

In considering the structural pathway for the transition, it is important to point out that the system may exhibit residual vibrational coherence in the product phase. However, because of the change in symmetry, the relevant vibration is renormalized to the edge of the Brillouin zone and becomes invisible to optical probing. This is in contrast with what has previously been observed in systems where the atomic structural symmetry is preserved despite large modulations in the electronic structure[24,25]. Secondly, despite the timescale coincidence with the Ag modes, the structural pathway connecting the two phases is not measured directly. It is possible that electronic excitation modifies the vibrational spectrum of the solid, which then may not be best described by the equilibrium normal modes. Finally, although the data supports a view Band-like picture for the semiconducting phase, it is quite possible that electronic correlations stabilize the low-T phase by the formation of singlets on the $V^{4+}$ dimers.

In summary, we have shown that ultrafast spectroscopy on the sub-vibrational timescale can be applied to resolve ambiguous cause-effect assignments across phase transitions in strongly correlated electron systems. Based on the ultrafast response of to photo-excitation, we conclude that that the atomic arrangement of the high-T unit cell is necessary for the formation of the metallic phase of $VO_2$, even if the correlated d band is highly depleted (hole-doped). This result is suggestive of significant Band-like character for the low-T insulator.

Acknowledgements: We gratefully acknowledge discussions with G. Sawatzky, J.B. Goodenough, Y. Tokura, G. Benedek and N. Mannella. We also wish to thank G. Haller, J. Ager and Q. Xu for their help in the processing of the films. Advice from G. Cerullo during the assembly of the non-collinear Optical Parametric Amplifier is here acknowledged.

This work was supported by the Director, Office of Science, Office of Basic Energy Sciences, Division of Materials Sciences, of the U.S. Department of Energy under Contract No. DE-AC03-76SF00098.

Corresponding author, email: ACavalleri@lbl.gov

**FIGURE CAPTIONS**

**Figure 1**

Structural and electronic phases of $VO_2$. The high-T phase is Rutile, with the $V^{4+}$ ions approximately at the center of $O^{2-}$ octhahedra (not shown). The low-T phase is derived after pairing and tilting along the c axis, i.e. it has a unit cell of doubled size and has distortion of the oxygen cages. The $3d^3s^2$ Vanadium atoms contribute four electrons to fill the valence band, leaving one electron in the conduction band. In the high-T phase, the bands close to the Fermi level are the V3d bands, composed by a purely d// band (oriented along the c-axis) and by 3d π–hybridized bands, mixed with p orbitals of the oxygen ligands. In the low-T phase, dimerization splits the d// band, with contributions from both structural and electronic correlations. Distortion of V-O bonds lifts the $3d_\pi$ band above the Fermi level.

**Figure 2**

Reflectivity/transmission responses of free standing $VO_2$ / $Si_3N_4$ structures for 100-fs excitation pulses at 790-nm wavelength and 50 mJ/cm². The pump-probe experiments were performed at near-normal incidence.

**Figure 3**

Pump-probe reflectivity experiments of the photo-induced phase transition. The experiments are performed in the non-etched structure with variable pulse durations between 1.5 ps and 15 fs, as measured at the sample position. White light was amplified in 1-mm, 32°-cut BBO, pumped with 400-nm pulses crossing the seed light at 3.5°. Pulse-compression in a pair of prisms was used to minimize the duration of the pump-probe autocorrelation at the sample position. The experiments were conducted using pulses of 100-nm bandwidth (FWHM) centered around 650-nm.

**Figure 4**

**Figure 4a:** Time-resolved evolution of the reflectivity, measured using a Ti:Sa oscillator emitting 40 fs pulses at 850-nm. The inset displays the oscillatory part of the signal, obtained by subtracting the background. **Figure 4b:** Unpolarized, continuous-wave Raman spectra (dashed curve), acquired in backscattering geometry using 532-nm cw excitation. The modes at 5.85 THz and 6.75 THz are fully symmetric $A_g$ modes, while the lower mode is of $B_g$ symmetry[26]. Continuous curve: Fourier transform of the time-resolved oscillations, revealing the coherent excitation of the $A_g$ modes only.

## Low-T       High-T

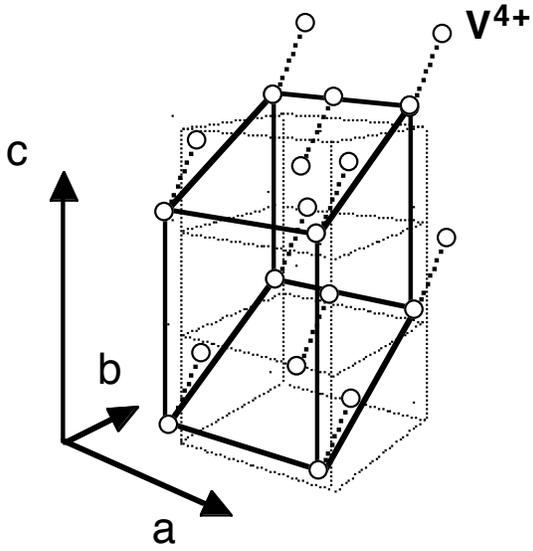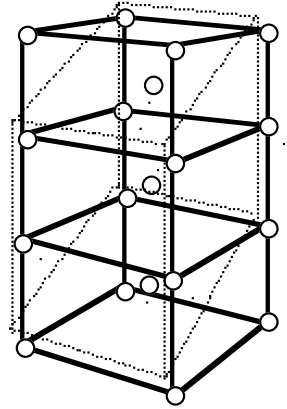

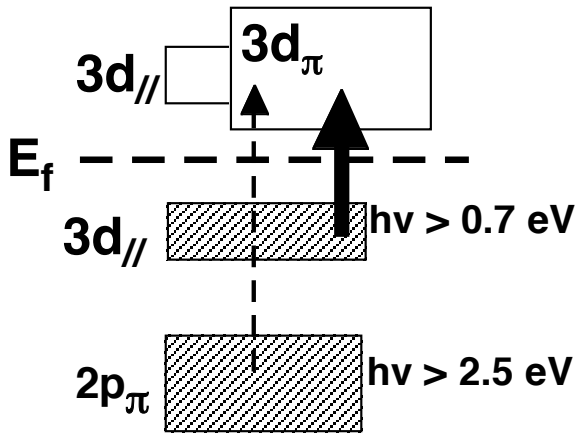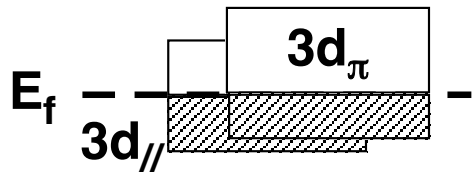

Figure 1

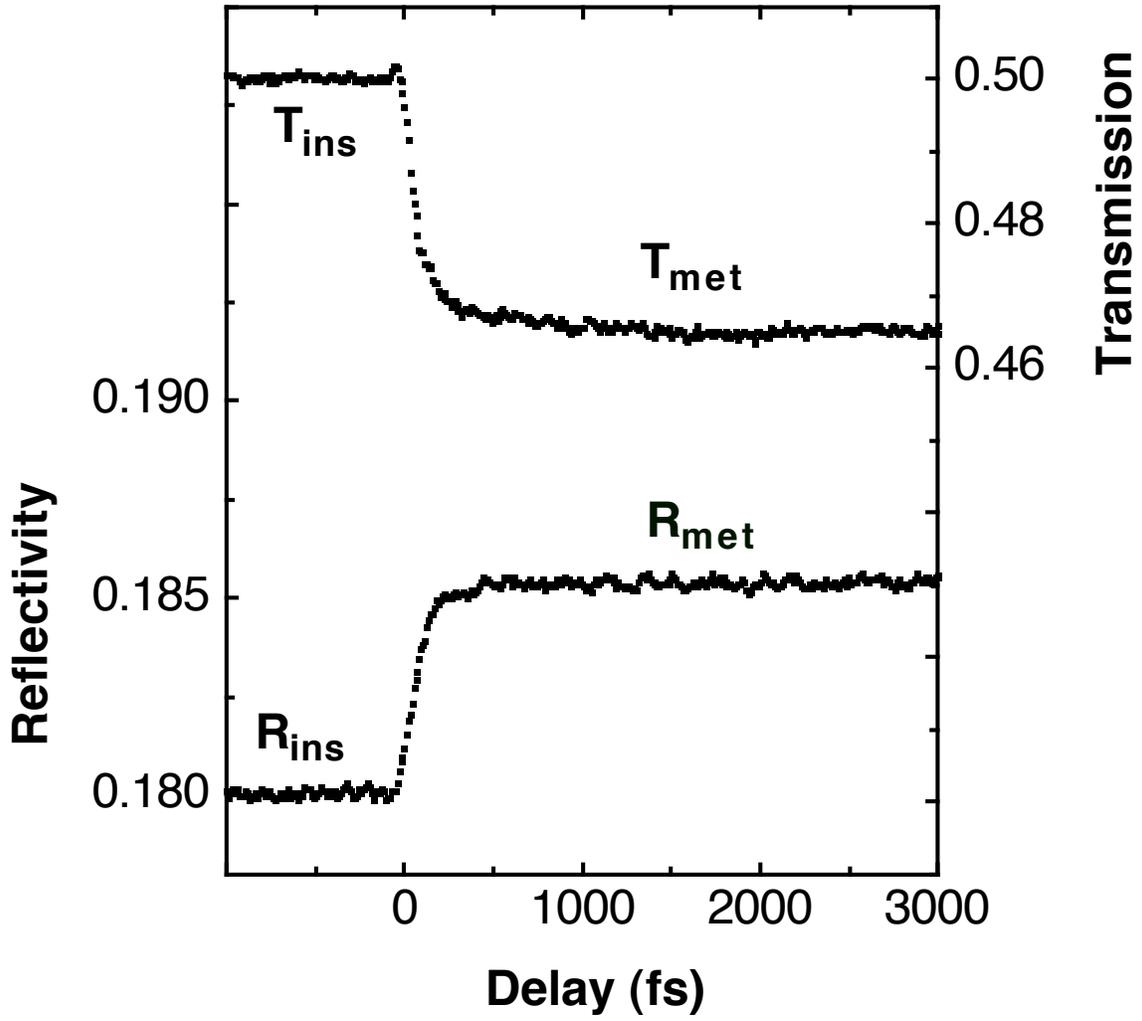

Figure 2

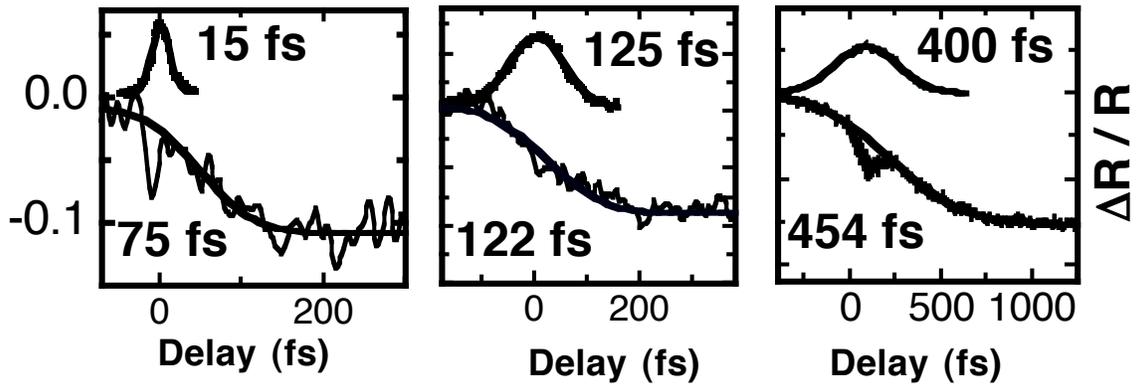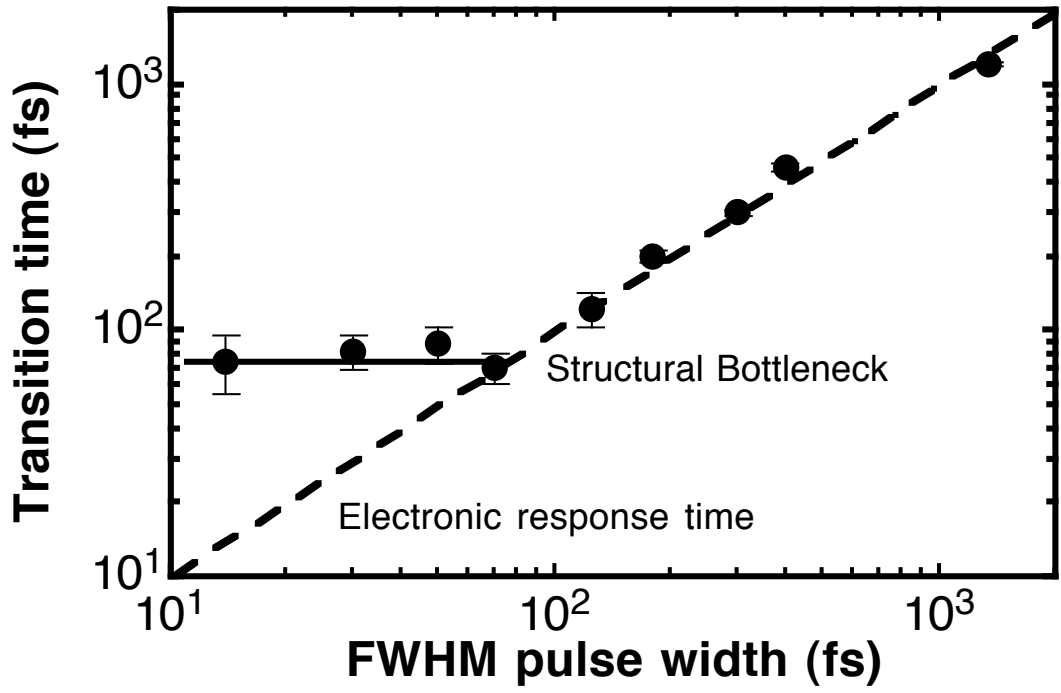

Figure 3

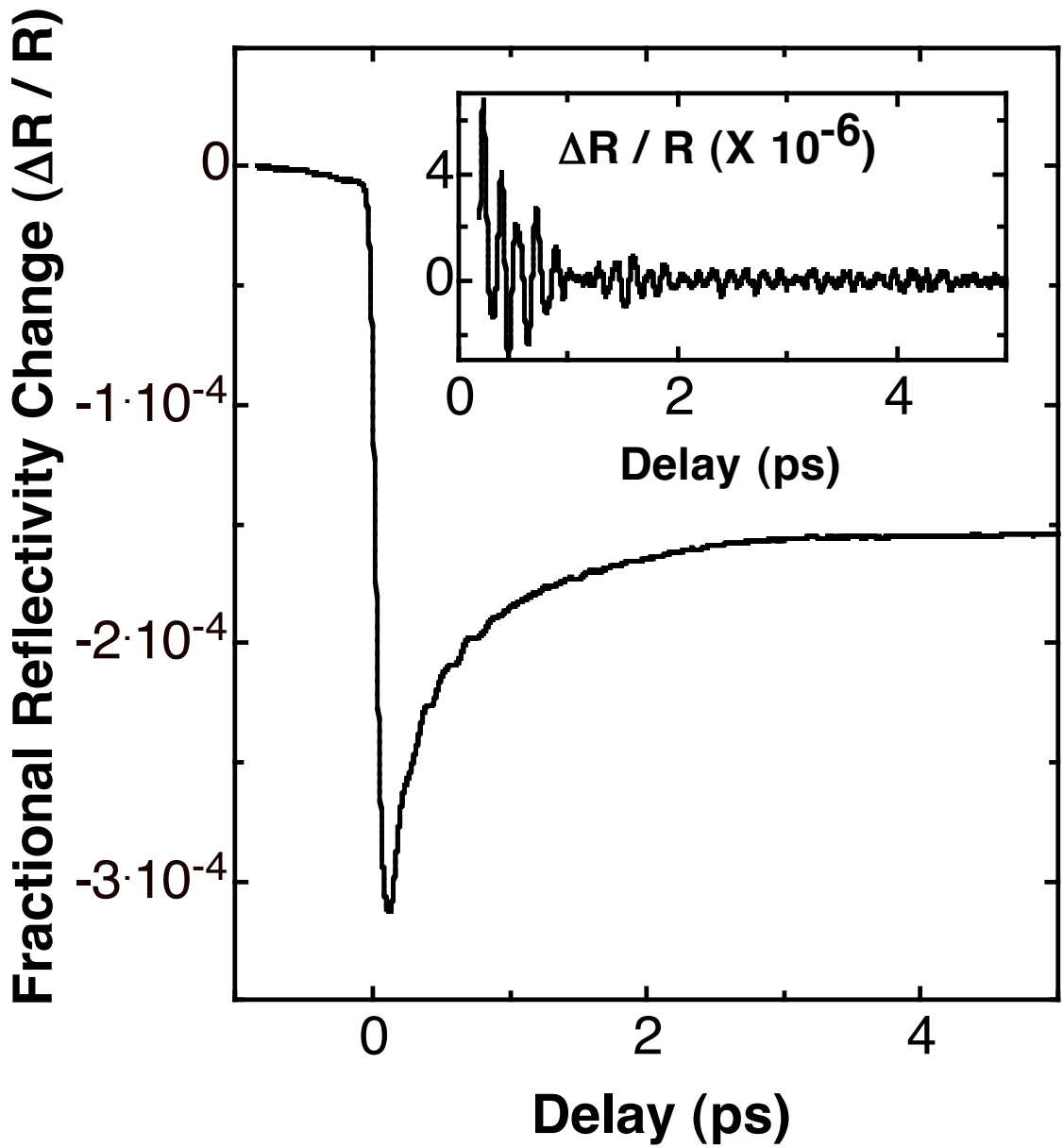

**Figure 4a**

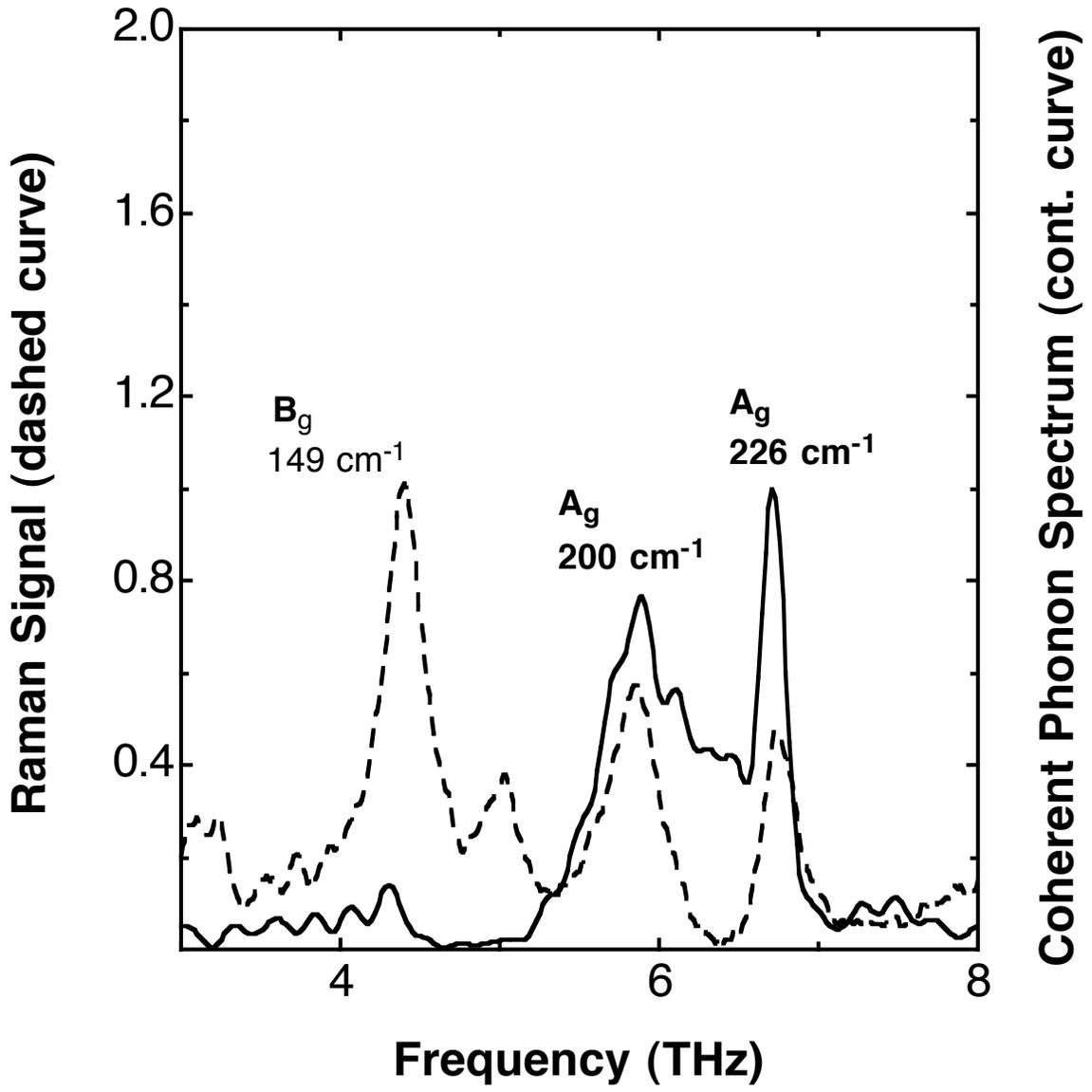

**Figure 4b**